\documentclass[showpacs,amsmath,amssymb,twocolumn,floatfix,prl]{revtex4}

\usepackage[dvips]{graphicx}

\begin{document}

\title{
On the Effective Charge of Hydrophobic Polyelectrolytes }

\author{A. Chepelianskii $^{(a)}$, F. Mohammad-Rafiee $^{(b,c)}$, E. Trizac $^{(d)}$ and E. Rapha\"el $^{(c)}$}
\affiliation{$^{(a)}$ Laboratoire de Physique des Solides, UMR CNRS 8502, B\^at. 510, Universit\'e Paris-Sud,
91405 Orsay, France } \affiliation{$^{(b)}$ Institute for Advanced Studies in Basic Sciences (IASBS), Zanjan
45195, P.O. Box 45195-1159, Iran. } \affiliation{$^{(c)}$ Laboratoire Physico-Chimie Th\'eorique, UMR CNRS
Gulliver 7083, ESPCI, 10 rue Vauquelin, 75005 Paris, France } \affiliation{$^{(d)}$
Universit\'e Paris-Sud, LPTMS, UMR CNRS 8626, 91405 Orsay, France }

\date{\today}

\begin{abstract}
In this paper we analyze the behavior of hydrophobic
polyelectrolytes. It has been proposed that this system adopts a
pearl-necklace structure reminiscent of the Rayleigh instability
of a charged droplet. Using a Poisson-Boltzmann approach, we
calculate the counterion distribution around a given pearl
assuming the latter to be penetrable for the counterions. This
allows us to calculate the effective electric charge of the pearl
as a function of the chemical charge. Our predictions are in
good agreement with the recent experimental measurements of the
effective charge by Essafi {\it et al.} (Europhys. Lett. {\bf 71}, 938 (2005)).
Our results allow to understand the large deviation from the Manning
law observed in these experiments.
\end{abstract}



\maketitle

\section{Introduction}

The study of polyelectrolytes has attracted an increased attention in the scientific community over the last
decades. This interest is motivated by technological applications including viscosity modifiers, or leak
protectors and by the hope that advances in this domain will allow to unravel the structure of complex
biological macromolecules. In these systems, the Coulomb interactions leads to many remarkable and
counterintuitive phenomena \cite{Manning1969,Oosawa,Kantor1992,Barrat,Holm}. A celebrated example is the
Manning-Oosawa counterion condensation. In his classical work \cite{Manning1969}, Manning showed that a charged
rod-like polymer can create such a strong attractive force on its counterions, that a finite fraction condenses
onto the polymer backbone. This condensation-phenomenon was also described by Oosawa within a two state model
\cite{Oosawa}. It leads to an effective decrease of the polymer charge, and the macroscopic properties of the
polyelectrolyte, like migration in an electrophoresis experiment, are not determined by its bare charge, but by
an effective charge that accounts for the Manning-Oosawa counterion condensation. It is now well-established
that counterion condensation is a fundamental phenomenon, and that it occurs in many important systems including
DNA in both its double-stranded and single-stranded form \cite{Bloomfield1999}. It was predicted in
\cite{Manning1969} that condensation occurs whenever the average distance $l$ between co-ions on the polymer
backbone is smaller than the Bjerrum length $\ell_B = q^2/(4 \pi \epsilon \epsilon_0 k_B T)$, where $q$ is the
co-ion charge, $k_B T$ the thermal energy and $\epsilon$ the (relative) dielectric constant of the solvent. This
condensation is expected to lead to an average charge density of $q / \ell_B$ on the polymer backbone. Since the
original prediction by Manning, important efforts have been devoted to a description of the Manning-Oosawa
condensation within the Poisson-Boltzmann theory and beyond
\cite{Philif,Shaughnessy,TriTell,Gonzalez-1995,Naji-2005}, establishing the influence of salt, the thickness of
the condensed counterion layer and the corrections induced by short range correlation.

While the conformation of many polyelectrolytes is well described by the rod-like model, many proteins organize
into complex self-assembled structures \cite{Alberts1998}. A challenging and important topic is the extent
to which the structural complexity of biological enzymes can be understood from simple physical models.
Polyelectrolytes with an hydrophobic backbone may provide an interesting system, that achieves a certain degree
of self-organization while the relevant interactions remain relatively simple. Indeed it has been predicted in a
seminal paper by Dobrynin and Rubinstein that hydrophobic polyelectrolytes should fold into an organized
pearl-necklace structure where regions of high and low monomer density coexist \cite{Dobrynin1996}. Therefore
both theoretical and experimental studies of the hydrophobic polyelectrolytes have shown a growing activity in
the past few years
\cite{Raphael,Raphael-91,Barrat,Carbajal2000,Dobrynin2001,Lee2002,Kyriy2002,Limbach-03,Holm,Dobrynin2005}.

The question of the validity of the Manning condensation model for hydrophobic polyelectrolytes has been
addressed experimentally by W. Essafi {\it et al.}  \cite{Damien-2005}. The authors have measured the effective
charge fraction of a highly charged hydrophobic polyelectrolyte (poly(styrene)-co-styrene sulphonate) by
osmotic-pressure and cryoscopy measurements. Their findings, which are recalled on Fig. 4, showed that the
measured effective charge is significantly smaller than that predicted by the Manning-Oosawa theory. The aim of
the present article is to provide a theoretical explanation of the counterion condensation in this system, where
the presence of hydrophobic interactions influences drastically the conformation of the polymer backbone. This
problem was first addressed theoretically by Dobrynin, and Rubinstein \cite{Dobrynin2001}, who determined the
phase diagram of a solution of hydrophobic polyelectrolytes as a function of solvent quality and polymer
concentration. However the question of the effective charge was not directly investigated by the authors.

The rest of the paper is organized as follows: In Sec. II the pearl-necklace model is reviewed briefly,
while the Poisson-Boltzmann theory of a hydrophobic globule permeable to counterions
is performed in Sec. III. The resulting effective charge is analyzed
in Sec. IV. Some aspects of the model are
discussed in Sec. V and finally, Sec. VI concludes the paper.

\section{II. Review of the pearl-necklace model}

Let us first recall for completeness the pearl-necklace theory of hydrophobic polyelectrolytes (for a more
complete review see \cite{Dobrynin2005}). The polyelectrolyte solution is parameterized by its degree of
polymerization $N$, its monomer size $b$, the charge fraction along the chain $f$, and the reduced temperature
$\tau \equiv 1 - \frac{\Theta}{T}$, where $\Theta$ and $T$ denote the theta temperature of the polyelectrolyte
and the temperature of the system, respectively. We note that in a bad solvent, the reduced temperature is
negative $\tau < 0$. We let $C$ be the average monomer concentration in the solution.

In a poor solvent, an uncharged polymer forms a globule in order to decrease its surface energy. In a similar
way, a drop of water adopts a spherical configuration in a hydrophobic environment.

To estimate the gyration radius $R_g$ of the polymer, we divide the polymer into smaller units, in such
a way that
inside each unit the thermal fluctuations dominate and the chain has Gaussian behavior. These units are usually
called thermal blobs in the literature and the typical radius of the blobs is denoted by $\xi_T$. It can be
shown that they contain about $1/\tau^2$ monomers, and have a typical size of $\xi_T \simeq b / |\tau|$. At
larger scales, the polymer tends to collapse onto itself in order to minimize its contact surface with the
liquid. This can happen by forming a dense packing of thermal blobs. A polymer of polymerization degree $N$ can
be split into $\tau^2 N$ thermal blobs and the volume occupied by the polymer is proportional to the number of
subunits. Therefore one can estimate the gyration radius of the polymer as
\begin{equation}
R^3_g \simeq \tau^2 N \xi_T^3 \simeq \frac{N b^3}{|\tau|}. \label{eq:Rg}
\end{equation}
The surface energy $E_S$ associated with this configuration is given by $k_B T$ times the number of thermal
blobs in contact with the solvent. This leads to
\begin{align}
\frac{E_S}{k_B T} \simeq \frac{\tau^2 R^2_g}{b^2}. \label{esurf}
\end{align}

\begin{figure}
\centering
\includegraphics[width=0.7\columnwidth]{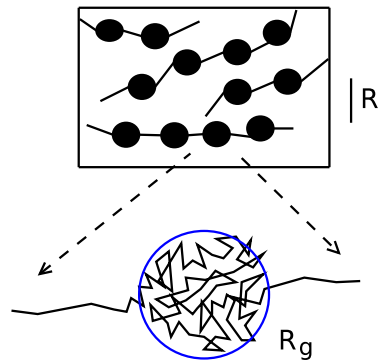} \\
\caption{Schematic drawing of the pearl necklace structure of hydrophobic polyelectrolytes: Inside the blue
(gray) circle the polymer backbone, represented by a continuous black line, is wrapped into a dense
configuration of typical radius $R_g$, that we call pearl or globule in the text. The inset shows on a larger
scale, that these pearls are connected by thin polymer strings, thus forming the pearl necklace structure. The
average distance between the pearls is $R$ (black vertical scale line). } \label{fig:Fig-pearl}
\end{figure}

Upon charging, the electrostatic repulsion sets in, which results in a change of the globule shape. When the
electrostatic repulsion energy becomes larger than the surface energy, the globule splits into several globules
of smaller size consisting of $N_g$ monomers. The typical size of these globules can be found from Eq.
(\ref{eq:Rg}) with the number of monomers $N_g$, which gives
\begin{equation}
R^3_g \simeq \frac{N_g b^3}{|\tau|}. \label{eq:Rg-Ng}
\end{equation}
This behavior is reminiscent of the Rayleigh instability of a charged droplet \cite{Rayleigh1882}. In this
state, the polyelectrolyte forms a sequence of globules that are connected by strings made of thermal blobs (see
Fig. \ref{fig:Fig-pearl}). In the literature, this conformation is known as the ``pearl-necklace'' structure.
The presence of counterions will screen the electrostatic repulsion. Therefore it is important to account for
their role explicitly in the  balance between the surface tension and the electrostatic repulsion that governs
the equilibrium structure of the necklace.

For simplicity we assume that the main effect of the counterions is to reduce the charge of the pearls. Indeed,
some counterions can be attracted inside the globules due to the attractive electrostatic forces. We assume that
the charged monomers of the polyelectrolyte have only one elementary charge $q$. In the absence of any counterion
condensation, the total electrostatic charge of a globule consisting of $N_g$ monomers is given by $q f N_g$. As
long as the counterions penetrate inside the globule, the effective charge of a globule is decreased and is
given by $q f_{eff} N_g$, where  $f_{eff}$ denotes the effective charge fraction. We can understand this
relation by noting that in the presence of counterion condensation the total charge of the pearl is the chemical
charge of the pearl minus the charge of the counterions inside it. Therefore, the electrostatic energy $E_{el}$
of a pearl can be estimated as
\begin{align}
\frac{E_{el}}{k_B T} \simeq \frac{ \ell_B \, (f_{eff} N_g)^2 }{R_g}, \label{eq:Eelec1}
\end{align}
where the Bjerrum length is defined as
\begin{equation}
\ell_B = \frac{q^2}{4 \pi \epsilon \epsilon_0 k_B T}, \label{eq:lB}
\end{equation}
where $\epsilon$ is the dielectric constant of the medium and $k_B T$ denotes the thermal fluctuation energy.
For example for water at room temperature ($T = 300 \; K $, $\epsilon = 80$) the value of the Bjerrum length is
$\ell_B \approx 0.7 {\rm nm}$. Using the relation between $R_g$ and the $N_g$ given in Eq. (\ref{eq:Rg-Ng}), the
electrostatic energy of a pearl is simplified to
\begin{equation}
\frac{E_{el}}{k_B T} \simeq |\tau|^2 \, f_{eff}^2 \, \frac{\ell_B R_g^5}{b^6}. \label{eq:Eelec2}
\end{equation}
In its equilibrium configuration the pearl-necklace tends to balance its electrostatic and surface energies
$E_{el} \simeq E_{S}$. Inserting the results of Eq. (\ref{esurf}) and Eq. (\ref{eq:Eelec2}) in this equality
leads to an expression of the globule radius $R_g$ as a function of the effective charge fraction $f_{eff}$:
\begin{align}
R_g \simeq b \left( \frac{b}{\ell_B} \right)^{1/3} \frac{1}{f_{eff}^{2/3}}.  \label{rgfeff}
\end{align}
We stress that this relation between the typical pearl size and the effective charge has been verified experimentally by D. Baigl {\it et al.}
in \cite{Baigl-2003} with an X-ray diffraction technique. This suggests that the hydrophobic polyelectrolytes studied in the experiment of W. Essafi actually
formed a pearl necklace structure.

\section{III. Screening of a globule in the Poisson-Boltzmann theory}

The problem of the effective charge of spherical microion-permeable
globules of size $R_g$ surrounded by their own counterions can
be solved in the mean-field approximation using the Poisson-Boltzmann theory. This problem was first studied
numerically and analytically by Wall and Berkowitz \cite{Wall-1957}.
It was shown that counterion condensation
around such a globule is possible (see e.g. Ref \cite{Bello} for a general discussion of the condensation
phenomenon).
In this approach, a charged globule is modelled as a sphere with radius $R_g$ and a uniform charge distribution
inside it. Therefore, the charge density of the globule is given by
\begin{equation}
 q \rho_0 \simeq q \frac{f N_g}{R_g^3}, \label{eq:rho0-1}
\end{equation}
where
$\rho_0$ denotes the mean density of charged monomers that are
distributed inside the globule. Using Eq. (\ref{eq:Rg-Ng}), $\rho_0$ can be simplified to
\begin{equation}
\rho_0 \simeq \frac{f |\tau| }{b^3}. \label{eq:rho0-2}
\end{equation}
In the solution, the mean monomer concentration is denoted by $C$. As far as the counterions are distributed
inside an elementary cell of radius $R$ (Wigner-Seitz approach), the average concentration of the counterions is given by
\begin{equation}
n_{av} = f C. \label{eq:n-1}
\end{equation}
Using the electro-neutrality condition, one can find a relation between the radius of the elementary cell,
$R$, and the density of the charged monomers inside the globule as
\begin{equation}
\rho_0 R_g^3 = n_{av} R^3. \label{eq:rho0-n}
\end{equation}
Assuming the spherical symmetry in the charge distribution, all the quantities such as the electrostatic
potential, the counterion concentration, {\it etc} depend only on the distance $r$ to the center of the globule.
Under the assumption of a Boltzmann-distribution, the concentration profile $n(r)$ of the counterions is related
to the electrostatic potential $\phi(r)$ as
\begin{equation}
n(r) = n_{av} e^{\frac{q \phi(r)}{k_B T}}. \label{eq:n-phi}
\end{equation}
Inserting this expression into the Poisson equation $\nabla^2 \phi = -\frac{1}{\epsilon \epsilon_0} (q \rho_0(r)
- q n(r))$ leads to the well-known Poisson-Boltzmann (PB) equation:
\begin{equation}
\nabla^2 \phi = \frac{1}{r^2} \frac{d}{dr} \left( r^2 \frac{d \phi}{dr}\right) = - \frac{q \rho_0(r)}{\epsilon
\epsilon_0} + \frac{q n_{av} }{\epsilon  \epsilon_0} e^{\frac{q \phi}{k_BT}}, \label{eq:PB1}
\end{equation}
where $\rho_0(r)$ is given by
\begin{eqnarray}
\rho_0(r)=\begin{cases}
\rho_0 \simeq  \frac{f N_g }{R_g^3} & r\leq R_g,\\
0 & r>R_g.
\end{cases} \label{eq:rho0}
\end{eqnarray}
For our system with spherical symmetry in the charge distribution, the electric field is zero at $r=0$.
Electroneutrality also demands a vanishing electric field at the boundary $r=R$,
so that the boundary conditions for the above PB equation read
\begin{eqnarray}
\frac{d \phi(r = 0)}{dr} = \frac{d \phi(r = R)}{dr} = 0. \label{eq:boundary}
\end{eqnarray}

In an elementary cell with the average counterion density $n_{av}$, the Debye screening length $\lambda_D$ is
given by
\begin{equation}
\frac{1}{\lambda_D^{2}} = 4 \pi \ell_B n_{av}. \label{eq:Debye}
\end{equation}
After defining the reduced electrostatic potential, $u \equiv q \phi / (k_BT)$, and $x \equiv r/ \lambda_D$, PB
equation can be written as
\begin{eqnarray}
\frac{d^2 u}{d x^2} + \frac{2}{x} \frac{d u}{d x} = e^{u(x)} - A(x), \;  \frac{d u(0)}{d x} = \frac{d u(X)}{d x}
= 0,  \label{eq:PB2}
\end{eqnarray}
where $X$ denotes $R/\lambda_D$ and $A(x)$ is defined as
\begin{equation}
A(x) \equiv \frac{\rho_0(x)}{n_{av}}. \label{eq:A}
\end{equation}
The radius of the globule in the dimensionless form is denoted by $x_g \equiv R_g/\lambda_D$. We will set $A$ as
the value of $A(x)$ inside the globule: $A(x) = A$ for $x \le x_g$. Using the aforementioned reduced variables
and the cell neutrality condition, Eq. (\ref{eq:rho0-n}), one can find the simple form of $A$ as
\begin{eqnarray}
A = \frac{\rho_0}{n_{av}} = \left(\frac{X}{x_g} \right)^3 \simeq \frac{|\tau|}{C b^3}, \label{eq:A-1}
\end{eqnarray}
where in writing the last term, the explicit forms of $\rho_0$, Eq. (\ref{eq:rho0-2}), and $n_{av}$, Eq.
(\ref{eq:n-1}), have been used. One can see that $A$ does not depend on the chemical charge $f$.

The fraction of counterions outside the globule, $P$, can be found as
\begin{equation}
P = \frac{\int_{r_g}^{R} n_{av} \, e^{q \phi(r)/k_BT } r^2 \, dr }{\int_{0}^{R_g} \rho_0 r^2 \, dr} =
\frac{\int_{x_g}^{X} e^u x^{2} d x}{ \int_{0}^{x_g} A x^{2} d x}, \label{eq:Integral-P}
\end{equation}
where in writing the last term, the reduced variables and Eq. (\ref{eq:A-1}) have been used. Using Eq.
(\ref{eq:PB2}) and integrating, leads to a simpler form of the above equation as
\begin{align}
P(x_g, A) = - \frac{3}{x_g A} \frac{d u(x_g)}{d x}. \label{eq:P}
\end{align}
As far as the penetrated counterions inside the globule reduce its charge, the effective charge of the globule
is proportional to the fraction of counterions outside the globule. Therefore, the effective charge of the
globule can be written as
\begin{align}
f_{eff} = P(x_g,A) f. \label{eq:feff}
\end{align}

It has been shown in \cite{Wall-1957} that the potential $u(x)$ defined by the boundary problem, Eq.
(\ref{eq:PB2}), is a decreasing function of $x$ and the initial value of the potential satisfies $e^{u(0)} \le
A$. Physically, this inequality signifies the absence of over-screening (inside the globule $q n(r) \le \rho_0$)
as expected in a mean-field theory \cite{Tr2000}. In order to estimate the lower limit of $e^{u(0)}$, we re-write Eq.
(\ref{eq:PB2}) as
\begin{eqnarray}
u(x) = u(0) + \int_0^{x} \left(y - \frac{y^2}{x} \right) \left[ e^{u(y)} - A(y) \right] dy. \label{eq:u-x}
\end{eqnarray}
Since $u(x)$ is a decreasing function, it may be shown that
\begin{equation}
u(x) = u(0) + \frac{1}{6} min(x,x_g)^2 \left[ e^{u(0)} - A \right],\label{eq:ux-u0}
\end{equation}
where $min(x,x_g)$ yields the smaller quantity. After inserting this result in the cell neutrality condition,
$\int_0^X x^2 \left[ e^{u(x)} - A(x) \right] dx$, we find that $e^{u(0)}$ satisfies
the following inequality relations:
\begin{equation}
1 - \frac{\ln{Z}}{Z} \le \frac{e^{u(0)}}{A} \le 1,  \label{ineq}
\end{equation}
where $Z$ is defined as $Z \equiv A x_g^2/6 > e$. Therefore, the above chain of inequalities implies that in the
limit where $A x_g^2 \gg 1$, we have $e^{u(0)} / A \rightarrow 1$.

\begin{figure}
\includegraphics[width=0.99\columnwidth]{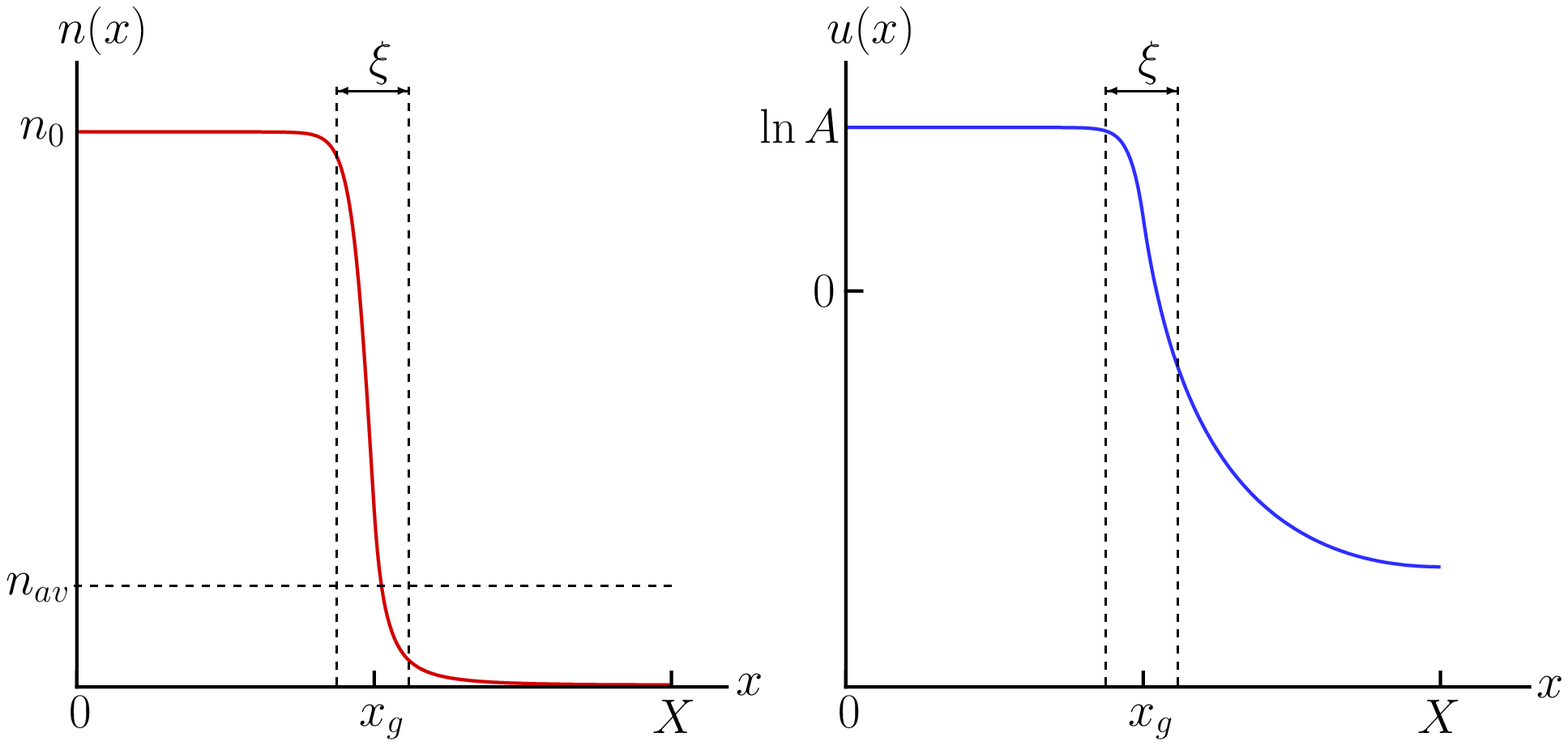} \\
\vspace*{-0.0cm} \caption{Typical behavior of counterion charge distribution $n(x)$ and effective potential
$u(x)$ in the cell. The dimensionless globule size is denoted by $x_g$ and the cell size is denoted by $X$.  }
\label{fig:n-u-x}
\end{figure}

The behavior of a typical solution $u(x)$ is displayed in Fig. \ref{fig:n-u-x}. It confirms that for large
values of $A x_g^2$, the counterion concentration at $x \simeq 0$ is very close to the concentration of charged
monomers inside the globule: $e^{u(x)} \simeq A$. As the value of $A x_g^2$ increases, the size of the neutral
region where $u(x) \approx \ln A$ grows until it becomes of the order of globule size $x_g$. Therefore, to keep
the system electrically neutral, the counterion concentration must fall down to values below $n_{av}$ outside
the globule.

The transition between these two regions occurs in a narrow layer of thickness $\xi$ on the boundary of the
globule, as shown in Fig. \ref{fig:n-u-x}. In order to estimate the behavior of $\xi$ in terms of physical
parameters in the problem, it is convenient to write PB equation for $x \gtrsim x_g$ in the following manner
\begin{eqnarray}
\frac{d^2 u}{dx^2} \left[ 1+ \frac{2}{x} \frac{ \frac{du}{dx}}{\frac{d^2u }{dx^2}} \right] = e^{u(x)}.
\label{eq:PB-xi}
\end{eqnarray}
We note that $d^2u / dx^2$ and $du/dx$ are of the order of $(\ln A)/ \xi^2$ and $(\ln A) / \xi$, respectively.
Putting these values in the above equation, we find
\begin{eqnarray}
\frac{\ln A}{\xi^2} \left[1 + 2\frac{\xi}{x_g} \right] \simeq A. \label{eq:xi-A}
\end{eqnarray}
We assume that we are in the regime where $\xi/x_g \ll 1$. Therefore, $\xi$ scales
\begin{eqnarray}
\frac{\ln A}{\xi^2} \simeq A \; \Longrightarrow \; \xi \simeq \frac{1}{\sqrt{A}}, \label{eq:scale-xi}
\end{eqnarray}
where we have neglected the logarithmic dependence on $A$. We note that
for consistency, the requirement $\xi \ll x_g$ also implies $A x_g^2 \gg 1$.

\begin{figure}
\vspace{0.1cm}
\includegraphics[width=0.8\columnwidth]{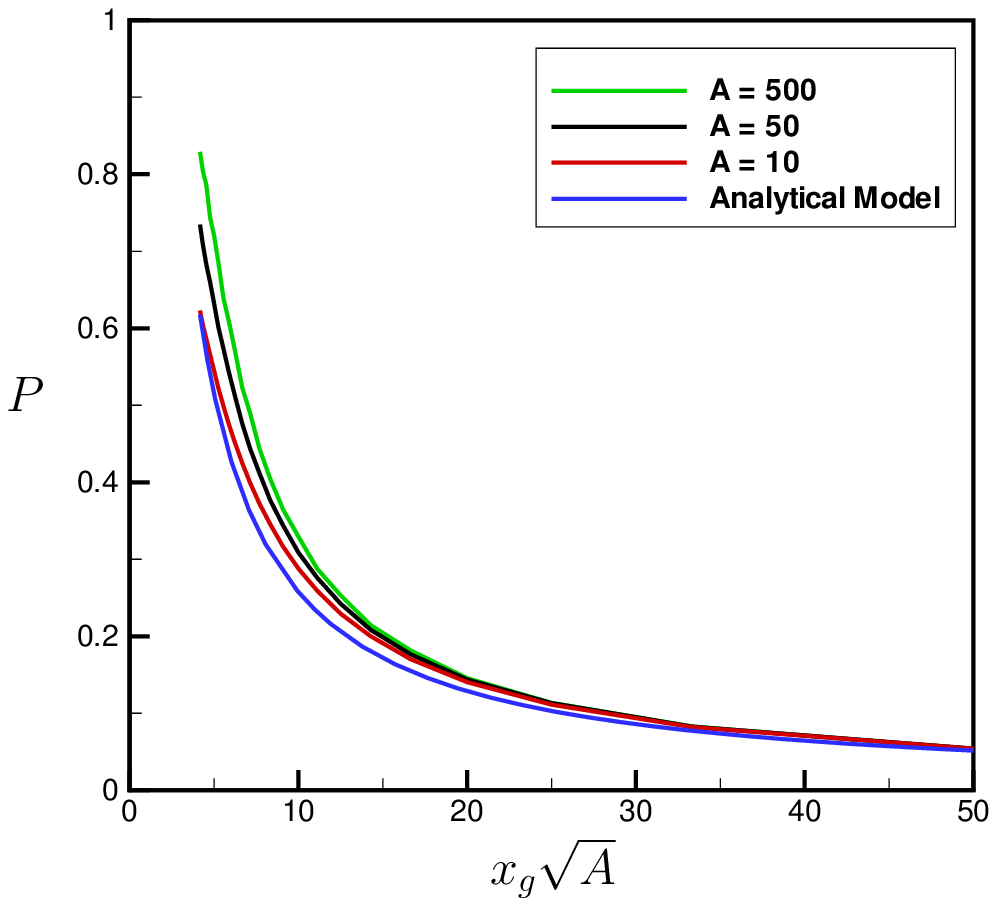} \\
\vspace*{-0.0cm} \caption{Dependence of $P$ on $x_g \sqrt{A}$ for different values of $A$. From top do bottom,
$A = 500, 50, 10$ (green, black, red curves respectively). The bottom (blue) curve is Eq.~(\ref{eq:P-A}).  }
\label{fig:P-A}
\end{figure}\

We are now in a position to estimate the counterion concentration outside the globule. Using Eq. (\ref{eq:P}) in the limit of $A
x_g^2 \gg 1$, the fraction of counterion outside the globule is found as
\begin{equation}
P(x_g,A) \simeq  \frac{1}{x_g A } \frac{u}{\xi} \simeq \frac{6}{\sqrt{2 e}} \frac{1}{x_g \sqrt{A}}.
\label{eq:P-A}
\end{equation}
The proportionality constant in Eq. (\ref{eq:P-A}) is calculated by ignoring the first derivative term
$\frac{1}{x} \frac{du}{dx}$ in Eq. (\ref{eq:PB2}). Fig. \ref{fig:P-A} shows that there is a very good agreement
between the exact and the analytical approximation results in the limit of $A x_g^2 \gg 1$ ($\xi \ll x_g$). We
also see that for a wide range of $A$ values, our analytical theory gives a good numerical approximation for $P$
as far as $P \lesssim 0.4$. For example for $A = 500$, the relative error of our approximation is below $20\%$
in this region. The exact numerical results were obtained using the method described in \cite{Wall-1957}.

\section{IV. the effective charge of a hydrophobic polyelectrolyte}

In the regime explored experimentally by W. Essafi {\it et al.} \cite{Damien-2005}, the value of the
dimensionless parameter $A$ can be estimated as follows. For $|\tau| \simeq 1$, monomer concentration $C = 0.1
\; {\rm Mol \; L^{-1}}$ and the bond length in the polymer $b = 0.25 \; {\rm nm}$, the expected value of $A
\simeq |\tau| / (C b^3)$ is $A \simeq 10^{3} \gg 1$. The value of $x_g$ depends on both the chemical and
effective charge fraction, $f$ and $f_{eff}$, as
\begin{eqnarray}
x_g = \frac{R_g}{\lambda_D} \simeq
 \frac{|\tau|^{1/2}}{A^{1/2}} \left(
\frac{\ell_B}{b} \right)^{1/6} \frac{f^{1/2}}{f_{eff}^{2/3}}.
\label{eq:xg}
\end{eqnarray}

\begin{figure}
\vspace{0.1cm}
\includegraphics[width=0.8\columnwidth]{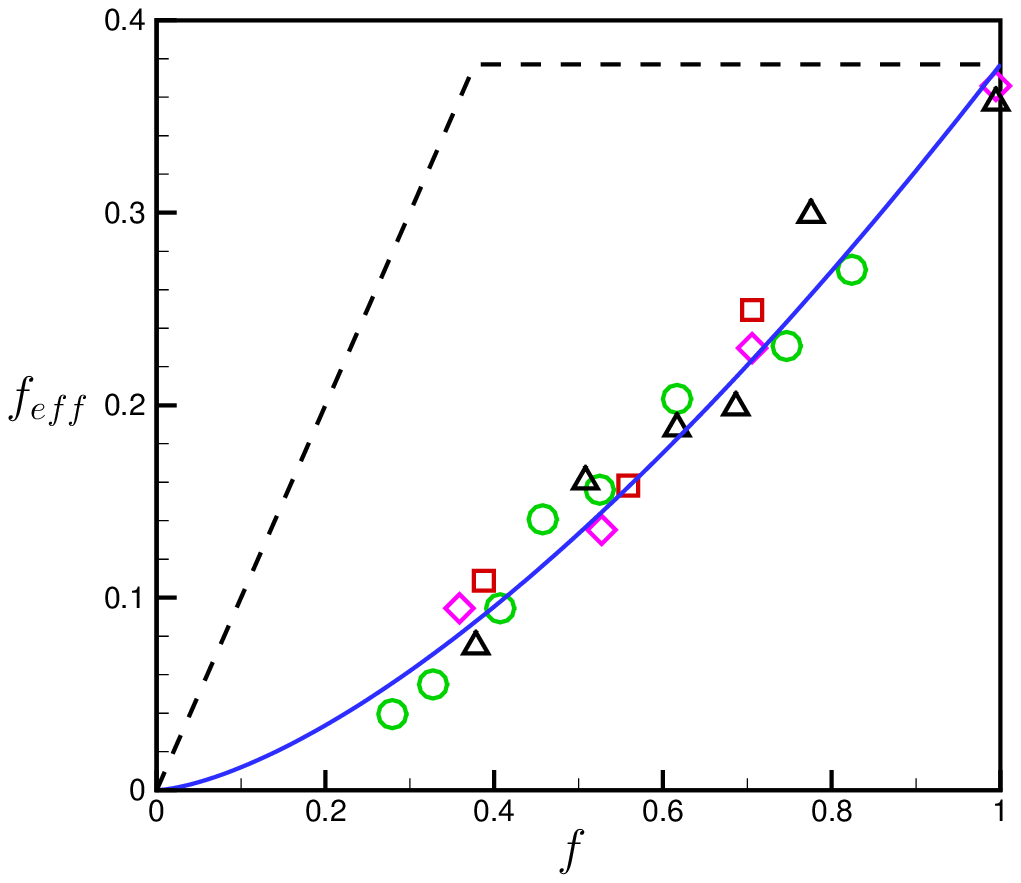} \\
\vspace*{-0.0cm} \caption{Effective charge fraction $f_{eff}$ versus the chemical charge fraction $f$. The
experimental points were obtained in \cite{Damien-2005}. The red squares correspond to $N=410$, green circles:
$N=930$, purple diamonds: $N=1320$, and black deltas: $N=2400$. The blue solid line corresponds to our
theoretical model Eq.~(\ref{eq:feff2}) with $\sqrt{ b / \left( |\tau|^3 \ell_B\right) } = 0.37$. The dashed line
corresponds to Manning's model.} \label{fig:feff}
\end{figure}

Using Eqs. (\ref{eq:feff}), (\ref{eq:P-A}), and (\ref{eq:xg}) the effective charge fraction $f_{eff}$ is found
as
\begin{eqnarray}
f_{eff} \simeq \sqrt{ \frac{b}{|\tau|^3 \ell_B} } f^{3/2}.
\label{eq:feff2}
\end{eqnarray}
This result predicts that the effective charge fraction $f_{eff}$ is proportional to $f^{3/2}$. We note that in
this regime the effective charge does not depend on the average monomer concentration $C$ and depends only on
intrinsic properties of the polymer. The scaling law of Eq. (\ref{eq:feff2}) and the experimental data of Fig.
4 of \cite{Damien-2005} are shown in Fig. \ref{fig:feff}. As one can see, there is a very good agreement
between the predicted behavior and the experimental data. We stress that only one free coefficient of order one
has been used to adjust the data. Thus, our theory can explain  the origin of the difference between the
effective charge predicted by the Manning-law recalled on Fig. \ref{fig:feff} and that observed in experiments.
Furthermore, using Eqs. (\ref{eq:Rg-Ng}), (\ref{rgfeff}), and (\ref{eq:feff2}) the globule radius $R_g$ and the
number of monomers inside the globule $N_g$ are found as
\begin{eqnarray}
R_g &\simeq& \frac{|\tau| b}{f}, \label{eq:rgeff} \\
N_g &\simeq& \frac{|\tau|^4}{f^3}. \label{eq:Ng-eff}
\end{eqnarray}

It is important to mention that in the experiments of \cite{Damien-2005}, only samples with relatively high
chemical charge fraction $f \ge 0.3$ were prepared, thereby limiting the range where our theory can be checked.
This is related to the difficulty to stabilize solutions of hydrophobic polyelectrolytes with low chemical
charge because the polyelectrolytes can form a macroscopic phase that is not soluble in the solvent. We expect
that the formation of a macroscopic phase can occur if the number of monomers inside a globule $N_g$ becomes
larger than the polymerization degree of the polymer $N$. In this case the polymer chains must stick together to
form globules of size $N_g \approx |\tau|^4/f^3 > N$, which may lead to form an entangled polymer network that
is not soluble in the solvent anymore. More detailed theoretical studies are needed on this problem. We note
that a detailed analysis of the possible phases and their stability range has been done in
\cite{Schiessel-1998}. As mentioned above, the dimensionless factors are of order one and if we set $N = 1000$
this condition for phase separation reads $f_{eff} < 1 / \sqrt{N} \simeq 0.03$. This result is in a reasonable
agreement with the results displayed in Fig. \ref{fig:feff}. It is worth to mention that in the experiments no
point could be obtained below this limit. We also emphasize that in our theory, when a stable pearl-necklace
structure forms, the effective charge depends on $N_g$ and not on polymerization degree $N$. This property has
been verified in the experiment, where $N$ has been varied from $N = 410$ to $N = 2400$ without apparent change
of the measured values of $f_{eff}$.

\section{V. Discussion}

In the above treatment, we have assumed that the polyelectrolyte chain in a dilute regime forms a necklace
structure in the solvent. Liao {\it et al.} \cite{Dobrynin-2006} have studied the necklace formation in
polyelectrolyte solutions using both theory and molecular dynamics simulations. They have shown that partially
charged chains form necklace-like structures of globules and strings in dilute solutions. For the dilute regime
the phase diagram of hydrophobic polyelectrolytes was obtained in \cite{Dobrynin-2006}. It has been shown that
when the effective charge of the chain is larger than a threshold $\sqrt{b |\tau| /(\ell_B N)}$, the necklace
structure is the dominant feature of the polyelectrolytes in a bad solvent. Using Eq. (\ref{eq:feff2}) and the
mentioned criterion, we find that for chains consisting of more than $1/ (|\tau|^2 f^3)$ monomers, the
necklace-structure is formed in the system. For the experimental condition explained in \cite{Damien-2005},
$|\tau| \simeq 1$ and $f > 0.2$, gives $1/ (|\tau|^2 f^3) \simeq 150$. All the chains that have been used in the
experiment \cite{Damien-2005} have more than 410 monomers on a chain, which means that our model considering
necklace structure for the hydrophobic polyelectrolyte in the solution is reasonable. We note that our approach,
does not allow to predict accurately the phase diagram of the polyelectrolyte chains. A consistent minimization
of the free energy requires to account correctly for the logarithmic dependence of the counterion entropic and
electrostatic energy as a function of the pearl radius \cite{Dobrynin-1999}. In our scaling law analysis, it is
not possible to follow the mentioned process. The origin of such logarithmic terms can be seen easily by
estimating the entropy of the counterions, since the condensed counterions explore only a phase volume of
$R_g^3$ out of the total volume. In our analysis this dependence is ignored because the available phase volume
is limited to the size of the Wigner-Seitz cell in a periodic system. Furthermore, the correlation induced
effects like the nonmonotonic dependence of the solution osmotic coefficient on the polymer concentration have
been observed in computer simulation analysis \cite{Dobrynin-2006}, which cannot be described in our model.

As we explained before, Eq. (\ref{eq:feff2}) is based on the validity of Eq. (\ref{eq:P-A}). It is justified
provided $ P \ll 1$ and our numerical calculations suggest that reasonable agreement is achieved for $P \lesssim
0.4$ for the experimental value of $A \simeq 10^3$. For the parameters used in Fig. \ref{fig:feff}, the
mentioned criterion is always satisfied. Furthermore, by placing the pearls inside neutral Wigner-Seitz cells,
we have ignored the effect of the interaction between neighboring pearls on the counterion distribution. However
the sharp decrease of the counterion concentration on the boundary of the globule (see Fig. \ref{fig:n-u-x})
suggests that these interactions should not affect significantly the counterion distribution. We have also
ignored the effect of the ions along the strings that connect adjacent pearls. This assumption can be checked by
estimating the fraction $s$ of the charged monomers present inside the pearls. It can be shown that
\begin{align}
s \simeq \frac{1}{1 + f_{eff} \sqrt{ \frac{\ell_B}{|\tau|^3 b} } }
\simeq \frac{1}{1 + f_{eff}},
\end{align}
where we have assumed that both the parameter $\sqrt{\frac{\ell_B}{|\tau|^3 b} }$ and intermediate scaling
constants are of order one. These assumptions are consistent with the parameters used in Fig. \ref{fig:feff}.
Our theory holds as long as $s \simeq 1$, that is when the effective charge $f_{eff}$ is be small. While this is
clearly the case in the range of small chemical charge $f$, the contribution of the strings may become important
when $f \simeq 1$. Physically we expect that around the strings, the counterions will follow the usual
Manning-condensation behavior. Therefore, the effect of the strings will be mainly to keep the effective charge
$f_{eff}$ below the Manning limit $b / \ell_B$. In Fig. \ref{fig:feff}, the effective charge reaches this limit
only at $f \simeq 1$; as a result the effect of the strings is not visible and our prediction holds even up to
$f \simeq 1$.

\begin{figure}
\vspace{0.5cm}
\includegraphics[width=0.8\columnwidth]{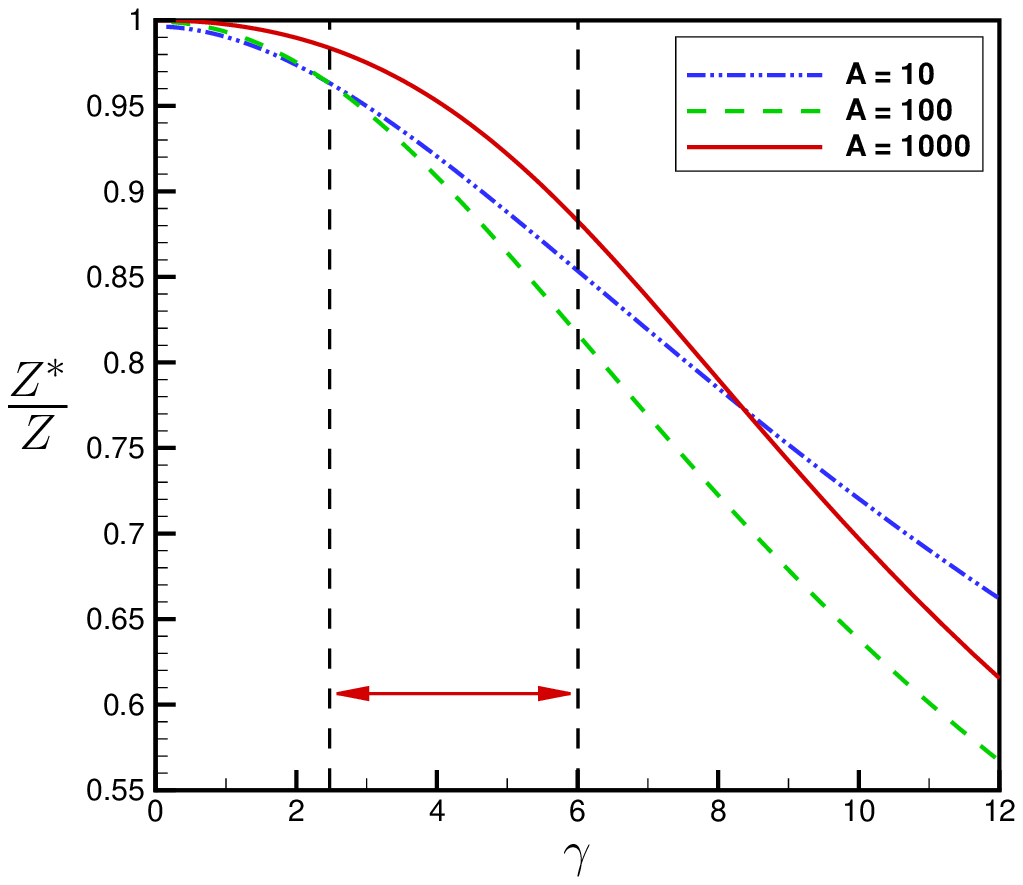} \\
\vspace*{-0.0cm} \caption{Ratio between the Alexander charge of the globule $q Z^*$ and the total charge inside
the globule $q Z = q f_{eff} N_g$ as a function of $\gamma = Z \ell_B/R_g$ for different values of $A$. ($A
\approx 10^3$ in the experimental conditions). The red arrow indicates the parameter range explored
experimentally by Essafi {\it et al.} estimated from Eq.~(\ref{eq:xixi}). }\label{fig:Alexander}
\end{figure}

Finally we have not taken into account additional counterion condensation outside the permeable globule. A
popular criterion for counterion condensation in this setting, was proposed by Alexander {\it et al.}
\cite{Alexander}. The renormalized charge $q Z^*$ of an impermeable globule of chemical charge $q Z$, is
determined from a linearization of the PB equation that ensures the best possible matching between the exact and
linearized solution at the boundary of the Wigner-Seitz cell. The dependence of the condensed fraction $Z^* / Z$
on the system parameters, is governed by the dimensionless parameter $\gamma = \left(Z \ell_B \right)/R_g$,
where $R_g$ is the globule radius \cite{Alexander,Belloni}. This parameter can be estimated as follows for the
case of our permeable pearl model. The charge fraction $Z$ of the pearl is given by $Z = f_{eff} N_g = f_{eff}
|\tau| R^3_g / b^3$, using Eqs. (\ref{eq:feff2}) and (\ref{eq:rgeff}) we find that
\begin{align}
\gamma = \frac{Z \ell_B}{R_g} \simeq \sqrt{ \frac{ |\tau|^3 \ell_B }{ b } } \frac{1}{\sqrt{f}} \simeq \frac{1}{
0.37 \sqrt{f} } \label{eq:xixi}
\end{align}
Here we used the numerical value for the coefficient $\sqrt{ |\tau|^3 \ell_B / b  }$ obtained from the fit to
the experimental data in Fig. \ref{fig:feff}. As a consequence, in the parameter regime explored experimentally
by W. Essafi {\it et al.}: $f \in (0.2, 1)$, we expect that $\gamma$ varies in the interval $\gamma \in (2.5,
6)$. We have calculated the ratio $Z^* / Z$ in this parameter range using the semi-analytical method proposed in
\cite{Trizac} and our numerical procedure. The obtained results are presented in Fig. \ref{fig:Alexander}, which
show that for $A \simeq 10^3$, the ratio $Z^* / Z$ is larger than $0.85$ when $\gamma \le 6$. This implies that
our expression for the effective charge is not renormalized significantly by condensation outside the globule.

\begin{figure}
\vspace{0.5cm}
\includegraphics[width=0.8\columnwidth]{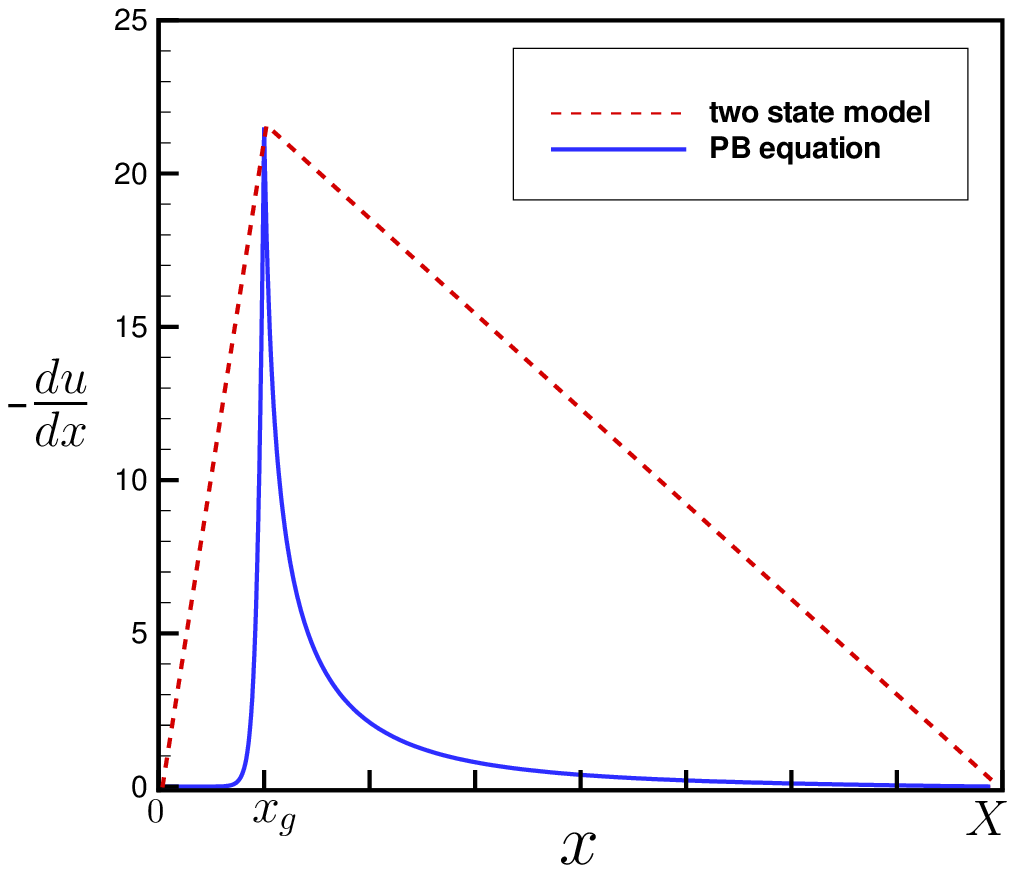} \\
\vspace*{-0.0cm} \caption{Typical behavior of the
dimensionless electric field $-\frac{du}{dx}$ using either the two state
model (dashed line) or our PB model (solid line). The solid line
corresponds to $x_g =1$ and $A = 500$.}\label{fig:E-x}
\end{figure}

It is interesting to compare our results to the results of \cite{Dobrynin2001}. Dobrynin and Rubinstein
considered for the first time the problem of counterion-condensation around an hydrophobic polyelectrolyte using
a two-state model. They determined the fraction $P$ by using trial counterion densities of the form $n(r) = (1 -
P) n_{av} \frac{R^3}{R_g^3}$ inside the globule (for $r < R_g$), and $n(r) = P n_{av} \frac{R^3}{R^3 - R_g^3}$
in the outer region. This family of density is parameterized only by the parameter $P$. Therefore by minimizing
the counterion free-energy density functional on this trial set, they could deduce an expression of $P$ as a
function of the system parameters \cite{Dobrynin-Note}. However for reasonable values of $|\tau|
\left(\frac{b}{\ell_B} \right)^{1/3} \simeq 1$, and for the experimental value of $A \simeq 10^3$, the value of
$f_{eff}$ predicted from the equations of ref. \cite{Dobrynin2001} is very close to $f$ in most of the parameter
range in contradiction with the experimental results of \cite{Damien-2005}. We attribute the difference between
our model and the results of \cite{Dobrynin2001} to the two state model used to estimate the fraction of
dissociated counterions $P$. Indeed in the two state model the charge density is constant in the two regions
inside and outside the globule. The Poisson equation then implies that in the two-state approximation, the graph
of the electric field ($- \frac{d u}{d x}$ in our dimensionless units) as a function of $x$ has a typical angle
shape for all values of $P$  as illustrated in Fig. \ref{fig:E-x}. In this figure, we have also compared this
approximation, to the exact numerical behavior of $- \frac{d u}{d x}$, for the typical parameters $A = 500, x_g
= 1$. Since the charged monomers at the center of the globule are neutralized by the counterions, the true
electric field distribution takes the form of a narrow peak centered at $x_g$. Because of its reduced family of
trial functions, the two-state model can not reproduce the true behavior of the electric field. However the
determination of the effective charge requires an accurate knowledge of the electric-field in the whole cell.
Therefore, we believe that the two state model is not accurate enough for the determination of the effective
charge. Indeed it was shown in \cite{Deserno-2001} that at least a three state model is necessary in the case of
a permeable droplet.

\section{VI. Conclusions}

In conclusion, we have developed a theory of counterion condensation around hydrophobic polyelectrolytes. Our
theory is based on the pearl-necklace model for the polyelectrolyte backbone. We assumed that the pearls are
permeable to the counterions, and use analytic results on the Poisson-Boltzmann equation to establish the
fraction of counterions condensed inside the pearls. It allows us to establish a power law dependence of the
effective charge $f_{eff}$ on the chemical charge $f$ as $f_{eff} \propto f^{3/2}$. This prediction is in very
good agreement with recent experimental results by W. Essafi {\it et al.} \cite{Damien-2005} and explains the
large deviation from the Manning law observed in these experiments. While our main results concern the effective
charge of hydrophobic polyelectrolytes, the scaling laws that we derived may also apply to other areas of
physics and chemistry where the Poisson-Boltzmann equation plays an important role.

\section{Acknowledgments}

We thank D. Baigl, M. Rubinstein, A. V. Dobrynin, and M. Maleki for fruitful discussions and precious remarks.
One of us, A. Chepelianskii, acknowledges the support of Ecole Normale Sup\'erieure de Paris.

\end{document}